\documentclass[5p,preprint]{elsarticle}

\usepackage{lineno}

\usepackage[english]{babel}
\usepackage{amsmath}        
\usepackage{amsfonts}
\usepackage{amssymb}        
\usepackage{bbm}         
\usepackage{booktabs}       
\usepackage{enumitem}
\usepackage{url}
\usepackage{hyperref}
\usepackage{multirow}
\usepackage{cleveref}
\usepackage{xspace}

\newlist{dashlist}{itemize}{1}
\setlist[dashlist]{label=--}

\newcommand{\EE}{\mathbb{E}}

\newcommand{\ATT}{\text{ATT}}
\newcommand{\indicator}{\mathbbm{1}}
\newcommand{\abs}[1]{\left| #1 \right|}
\newcommand{\floor}[1]{\left\lfloor #1 \right\rfloor}
\newcommand{\COtwo}{CO\textsubscript{2}\xspace}
\newcommand{\COtwoeq}{CO\textsubscript{2}eq\xspace}

\usepackage[dvipsnames]{xcolor}
\usepackage[normalem]{ulem}

\begin{document}

\begin{frontmatter}
    \title{Impact of Residential Retrofits on Gas and Electricity Consumption in France}
    \author[hw]{Charly Andral\corref{mycorrespondingauthor}}
    \cortext[mycorrespondingauthor]{Corresponding author}
    \ead{charly.an@hellowatt.fr}

    \author[hw]{Laetitia Leduc}
    \ead{laetitia.leduc.publications@proton.me}

    \author[hw]{Guillaume Matheron}
    \ead{guillaume_pub@matheron.eu}

    \author[hw]{Yukihide Nakada}
    \ead{publications@yukihi.de}

    \address[hw]{Hello Watt, 48 rue René Clair 75018 Paris, France}

    \begin{abstract}
        This study examines the impact of residential energy retrofits on household energy consumption in France using smart meter data from nearly 2,500 Hello Watt users, using a two-period difference-in-differences design.

        The dataset combines daily electricity and gas consumption collected through smart meters, hourly temperatures from Météo France, and user-declared home and retrofit information.
        As a control, we use a group composed of homes of Hello Watt users that are similar to the treated homes, but did not undergo any renovations.

        The average treatment effect on the treated is estimated with the estimator of Sant'Anna \& Zhao (2020). Estimates are reported by energy source (electricity vs.\ gas) and by retrofit type. The retrofit measures considered are limited to single interventions: wall insulation, attic insulation, floor insulation, installation of an air-to-air heat pump, or installation of an air-to-water heat pump. A comprehensive retrofit is defined separately as the simultaneous implementation of at least two of these measures.

        Our results show that insulation works cause a significant decrease in both electricity and gas consumption (3\% to 13\% and 5\% to 16\% respectively, depending on the retrofit type). We also estimate the reduction on the heating consumption only (7\% to 27\% for electrical heating and 7\% to 19\% for gas heating).
        We also study retrofits that consist in replacing a gas boiler with an air-to-water heat pump, resulting in a cut of 85\% in carbon emissions.
    \end{abstract}

\end{frontmatter}

\section{Introduction}
Hello Watt is a company that allows individuals to take control of their energy consumption through a free app that analyses their load curve as reported by smart meters.

Through disaggregation of the consumption curve\cite{culiereBayesianModelElectrical2020c,belikovDomainKnowledgeAids2022,brillandOccupancyDetectionBased2023}, users are guided towards the most relevant options to reduce their consumption, which can include services that Hello Watt offers, such as insulation or replacement of their heating system.

As of September 2025, close to one million homes have shared data consumption data with Hello Watt, which comes with user-provided characteristics about homes such as heating system and renovations.
Crucially, this data includes any previous retrofits, giving Hello Watt a unique perspective on consumption patterns before and after retrofits.

Home retrofits are a valuable tool for residential homes to decrease their long-term carbon footprint, and the French government's subsidies over the years are evidence of the importance of home retrofits in the national conversation about energy efficiency and decarbonization.
Home comfort has also recently developed into an important part of the conversation, as extreme weather in France takes an increasing toll on homes which are not built to regulate inside temperatures; our previous study which demonstrated that residents with well-insulated homes do not consume much less than a less well-insulated home \cite{matheronComparerDPEConsommation2023} suggests that increasing insulation may improve comfort in addition to reducing consumption.
Hello Watt, both through its app and renovation department, accelerates the adoption of home retrofits.

Quantitative analysis of the impact of retrofitting, such as this study and the recent study by INSEE \citep*{babamoussaEffetsLisolationThermique}, is crucial for informed policy-making regarding subsidies and incentives.
An analysis requires controlling for a large number of variables, such as different coverage periods, people making several retrofits at different times or using concurrently different kinds of heating systems, etc. \cite{economeePourquoiTravauxRenovation2025}.
This is only possible using a very large initial sample of homes, which Hello Watt is able to provide.

In this paper, we analyze the impact of a wide variety of home retrofits through a difference-in-differences framework, controlling for covariates such as home surface, house age, and geographical location. More precisely, the primary objective of this study is to estimate the causal effect of retrofitting on household energy consumption, which we formalized with the concept of the \textit{average treatment effect on the treated} (ATT) from causal inference \citep*{bakerDifferenceinDifferencesDesignsPractitioners2025}.

A novel contribution of this study is a robust measurement of the impact of installing heat pumps on home energy consumption, and access to more precise home characteristics such as the presence of swimming pools, electric vehicles, and the type of heating and water heating.

We use data from several thousand French users of the Hello Watt app \citep*{wattSuivreSaConsommation} who declared having done a retrofit work on their home between 2020 and 2024.
We separate homes by the main energy type used for heating, namely electricity and natural gas, and perform the study on each group. A third smaller group is composed of homes that switch from gas heating to air-to-water heat pumps.

The paper is organized as follows:
\begin{dashlist}
    \item In Section \ref{sec:related_work}, we compare our methodology to related works.
    \item In Section \ref{sec:data} we present the data used in this analysis. They are split into two main categories: time series data (weather, energy consumption) and declarative data coming directly from Hello Watt users (retrofit and home characteristics).
    \item In Section \ref{sec:methodology}, we describe the methodology used to estimate the ATT of the retrofit work on energy consumption.
    We use a 2 by 2 difference-in-differences (DiD) schema, with two time periods (before and after the retrofit work) and two groups (treated and control). The estimation of the ATT is based on the doubly-robust estimator described by \cite*{santannaDoublyRobustDifferenceindifferences2020}. The control group is selected to match the control group as closely as possible, using $k{:}1$ nearest neighbors matching on home characteristics.
    \item We present the main results of the ATT estimation in Section \ref{sec:results}. We found that retrofitting homes reduces their energy consumption, with the effect varying by the type of retrofit and the energy source used for heating.
    \item Finally, we discuss the results and limitations of our analysis in Section \ref{sec:discussion}.
\end{dashlist}

\section{Related Work}
\label{sec:related_work}

\paragraph{Methodology}
The evaluation of retrofit policies and household energy efficiency interventions has long relied on econometric methods designed to isolate causal impacts from observational data.
A large body of research has used quasi-experimental approaches such as matching estimators, instrumental variables, and difference-in-differences (DiD) to assess the effectiveness of energy efficiency programs \citep[e.g.][]{allcottSocialNormsEnergy2011}.

The canonical DiD framework, rooted in labor economics \citep{ashenfelterUsingLongitudinalStructure1984}, has been widely applied to policy evaluation, including in the energy domain.
Recent methodological contributions have highlighted important limitations of the two-way fixed effects (TWFE) estimator commonly used in DiD settings, particularly in the presence of heterogeneous treatment effects or staggered adoption.
To address these concerns, alternative estimators have been developed that provide more robust identification of the average treatment effect on the treated (ATT).

Among these, the doubly robust DiD estimator proposed by \cite{santannaDoublyRobustDifferenceindifferences2020} has emerged as a flexible and efficient tool for causal inference with panel data.
Applications in economics and social sciences have demonstrated its advantages in settings with covariate imbalance and treatment effect heterogeneity.

Our methodological approach is therefore consistent with this line of work.

\paragraph{Empirical results}

Our paper has similarities with a study published by the French INSEE\citep*{babamoussaEffetsLisolationThermique} in 2025, where consumption data is extracted from smart meters and retrofits are estimated using subsidies requests. This study also uses the DiD framework and reports average decreases in consumption of about $5.4\%$ (resp.\ $8.9\%$) in electrically (resp.\ gas) heated homes, with larger effects among high-consumption households. Contrary to our work, this study does not include results on the installation of heat pumps.
The French government agency ADEME recently published two studies focused on heat pumps: one for air-to-air heat pumps \citep{dupretEtudeConsommationsPAC2025} and one for air-to-water heat pumps \citep{ademePerformancePompesChaleur2025}. The first, based on 88 homes, reports a reduction in heating consumption by a factor of 2 for air-to-air heat pumps. The second, based on 90 homes, assesses the real-world performance of air-to-water heat pumps and finds a COP of 2.9.

Having a direct relationship with users allows Hello Watt to collect more information about the characteristics of a home that may impact its consumption, such as the presence of a swimming pool or an electric vehicle. This leads to more robustness to bias, and is a strong confirmation of previous results, with the addition of results on the installation of heat pumps.

Internationally, billing-data evaluations summarized in \citep{giandomenicoSystematicReviewEnergy2022} find average realized savings of about $7\%$ across nearly 140{,}000 households.
This order of magnitude is consistent with our results and highlights that realized savings are systematically lower than engineering expectations.
By contrast, building-physics case studies and simulation exercises often report much larger reductions when focusing on space-heating demand or deep renovation packages.
For instance, a Swedish multi-family renovation combining envelope and ventilation measures documented a reduction of about $40\%$ in heating demand \citep{lafleurMeasuredPredictedEnergy2017}, while simulations for Southern Europe suggest primary-energy savings up to $60\%$ depending on climate and renovation depth \citep{wangImpactEnergyRenovation2022}.
Experimental evidence from the U.S. Weatherization Assistance Program similarly shows realized savings below projected levels, underlining a persistent performance gap \citep{fowlieEnergyEfficiencyInvestments2018}.

\section{Data}
\label{sec:data}
The data used in this analysis can be split into two main categories: declarative data (retrofit and home characteristics) coming directly from Hello Watt users, and time series (weather and energy consumption) that we collect from weather and energy providers, respectively.

\subsection{Declarative data from Hello Watt users}

When installing and configuring the Hello Watt app, users declare their home characteristics such as inhabitable surface, number of inhabitants, presence of a swimming pool, and whether any retrofit works have been done on their home, along with the nature and date of the work.

As this is purely declarative data, it suffers from several potential issues:
\begin{dashlist}[noitemsep, topsep=3pt]
    \item some homes have partial information because some options have been added to the user interface after their users joined the app;
    \item users may not declare their retrofit and home characteristics, or only partially declare it;
    \item users may declare inexact or incorrect characteristics.
\end{dashlist}

\subsubsection{Home characteristics}
\label{sec:methodology:home_metadata}

We use self-reported home characteristics in order to select the control panel, and as confounding variables for the DiD estimator.

We detail below the home characteristics used in this study, as well as the values they can take:
\begin{dashlist}[noitemsep, topsep=3pt]
    \item home surface (in m$^2$)
    \item number of inhabitants (\{1, 2, 3, 4 or more\})
    \item number of floors (\{1, 2, 3 or more\})
    \item house age (\{before 1919, 1919-1945, 1946-1970, 1971-1990, 1991-2005, after 2005, unknown\})
    \item swimming pool (\{yes, no\})
    \item electric vehicle (\{yes, no\})
    \item heating system (\{heat pump, electric (except heat pump), gas\}\footnote{The variable can take other values, but they are rare and are excluded from the analysis as we are interested in the main heating system.})
    \item secondary heating system (\{heat pump, electric (except heat pump), gas, fireplace, pellet stove, wood stove, oil, other\})
    \item water heating system (\{electric, gas, other\})
    \item latitude and longitude
\end{dashlist}

We present in \Cref{tab:house_characteristics_elec} and \Cref{tab:house_characteristics_gas} the distribution of these home characteristics in the control and treated groups for electrical and gas heating, respectively.

\subsubsection{Retrofit data}
Hello Watt users can declare their retrofit works in the app. The works considered in this analysis are:
\begin{dashlist}[noitemsep, topsep=3pt]
    \item attic insulation
    \item wall insulation
    \item basement insulation
    \item window replacement
    \item heat pump installation
\end{dashlist}
For each retrofit, users declare the completion date, which is used to define the pre- and post-work periods in the difference-in-differences analysis.

Performing several retrofit measures at once has implications for subsidies in France as outlined in~\citep{MaPrimeRenovPourRenovation}, therefore in our analysis we also include a composite retrofit measure, which we call `comprehensive'.

\subsection{Consumption data}
\label{sec:energy:data}

The main objective of the Hello Watt app is to help users understand and optimize their energy usage and spending.
As such, users are given the option to input their smart meter identifiers (Linky for electricity and/or Gazpar for gas) and consent to Hello Watt collecting and analyzing their consumption patterns.
The data is then fetched from the APIs of Enedis~\cite{CompteurLinkyLallie} for electricity and GRDF~\cite{ToutSavoirCompteur} for gas.

The finest available resolution is usually 30 minutes for electricity and 1 day for gas.
As we will aggregate the data to a monthly resolution, we only use daily energy consumption for both energies, expressed in kWh.
An important clarification is that in case of gas, the value returned by the meter is the higher heating value (HHV, also known as gross calorific value, GCV) of the gas.
This is the value that is used to calculate the energy bill.
Thus, all gas consumptions in this article are expressed in kWh HHV, even if not explicitly mentioned.

Consumption data is collected at the individual user level.
For some users, both gas and electricity records are available, while for others only one source is present.
Consequently, the number of observations differs across energies and across users, and coverage does not always align over identical time periods.
This heterogeneity in data availability has direct implications for the design of the difference-in-differences analysis.
In particular, it motivates our choice to define time relative to the retrofit completion date, rather than in absolute calendar dates. This choice is further detailed in Section~\ref{sec:methodology}.

\subsection{Weather data}

We use weather data provided by Météo France~\cite{DonneesPubliquesMeteoFrance}, available with an Etalab Open License.
The only weather variable used in this analysis is the hourly temperature, which is then averaged per day to obtain a daily temperature.
We use the data from 521 weather stations, selected based on their availability between 2015 and 2024.

From this daily temperature, we compute an estimate of heating requirements using heating degree days (HDD)~\cite{dayDegreeDaysTheory2006}, which are obtained for each day $d$ as:
\begin{equation}
    \text{HDD}_d = \max(0, T_d - T_{\text{ref}})
\end{equation}
where $T_d$ is the average temperature on that day and $T_{\text{ref}}$ is a fixed reference temperature.
In this study, we use $T_{\text{ref}} = 15$°C, which is also the one selected by \cite{bruguetWeatherEffectsEnergy2025}.

From this daily HDD, we define $\text{HDD}_{m,y}$ for a given month $m$ and year $y$ as the sum of the $\text{HDD}_d$ for each day $d$ in that month and year.

Finally, we compute $\text{HDD}_{m}^\text{ref}$ using historical weather data as the average $\text{HDD}_{m,y}$ over the years 2015 to 2024.

\subsection{Correction of consumption data for weather effects}\label{sec:methodology:weather_correction}

\label{sec:methodology:heating_degree_days_correction}
As mentioned in the introduction, we use only two time periods for the difference in differences schema.
We thus need to aggregate the energy consumption data to two time periods: before and after the retrofit work.
However, we must account for the fact that weather can evolve between the two periods; for instance, the winter before the retrofit may be significantly colder than the winter following.

This is partially accounted for using geographical proximity as a criterion for the construction of the control group.
We also directly de-bias the consumption data using HDDs, as follows.

We define the monthly consumption $C_{m,y}$ for a given month $m$ and year $y$ as the average consumption of the home during that month and year.
In order to separate thermosensitive (heating) and non-thermosensitive (base) consumption, we compute $C^{\text{base}}$  as the mean consumption of the home during May, June and September for electricity consumption and during May, June, July, August and September for gas consumption.
This corresponds to the periods when no heating or air conditioning is used.
This is computed separately for each energy for each time period, thus allowing a different base for each energy and each time period.

We then define the corrected consumption for a given month $m$ and year $y$ as:
\begin{equation}
    C_{m,y}^{\text{corrected}} =
    \begin{cases}
        \tilde{C}^{\text{corrected}}_{m,y} & \text{if } \text{HDD}_{m,y} > 10 \text{ and } C_{m,y} > C^{\text{base}} \\
        C_{m,y}                            & \text{otherwise},
    \end{cases}
\end{equation}
where
\begin{equation}
    \tilde{C}^{\text{corrected}}_{m,y} = (C_{m,y} - C^{\text{base}}) \cdot \frac{\text{HDD}_{m}^{\text{ref}}}{\text{HDD}_{m,y}} + C^{\text{base}}.
\end{equation}

We also use this decomposition to estimate the heating part of the consumption, defined as
\begin{equation}
    \label{eq:heating_consumption}
    C_{m,y}^{\text{heating}} = \max(0,C_{m,y}^{\text{corrected}} - C^{\text{base}}),
\end{equation}
and the base part as
\begin{equation}
    \label{eq:base_consumption}
    C_{m,y}^{\text{base}} = C_{m,y}^{\text{corrected}} - C_{m,y}^{\text{heating}}.
\end{equation}
Those estimates are used to isolate the impact of the retrofit on the heating part of the consumption.

We choose not to add weather to the DiD estimator as a confounding variable, since the doubly-robust estimator is not compatible with time-dependent covariates.

\subsection{Data selection}
\label{sec:data_selection}
Once the different types of data are collected and merged, we select the homes that are relevant for the analysis. Homes heated with electricity and those heated with gas are treated separately.

Among all homes which have declared a retrofit, we select those where:
\
\paragraph{There is at least one year of energy data before and after the retrofit}
All the homes that have declared a retrofit work must have at least one year of energy consumption data before and after the retrofit work.

\paragraph{The retrofit work is done alone}
The retrofit work must be done alone and not in combination with other works.
A separate `comprehensive' work is used to describe homes that have performed two or more individual retrofits simultaneously.

\paragraph{Some heating is detected before and after the retrofit}
Homes must have electrical (resp. gas) heating consumption before and after the retrofit work.
This filter is based on an analysis of the consumption curves, and prevents the inclusion of homes that changed their primary heating method without declaring it in the app.
If both elec and gas consumption are available for a home, we also check that the other energy is not used for heating using again the energy consumption and HDD data.
This avoids having to compare electrical and gas consumptions.

\section{Methodology}
\label{sec:methodology}

The objective of this study is to quantify the causal impact of residential retrofit measures on household energy consumption.
Since retrofits are not randomly assigned, a naive comparison of treated and random untreated homes would conflate the effect of the intervention with differences in household characteristics and energy usage patterns.

To overcome this challenge, we adopt an econometric framework based on the average treatment effect on the treated (ATT) and identify it using a difference-in-differences (DiD) design.
Bias is further reduced by using a control group which is composed of homes that have similar characteristics to the treated group.

\subsection{Estimating the impact of retrofit measures on energy consumption}

This approach compares the change in energy consumption before and after retrofit for treated homes to the corresponding change in a carefully selected control group of non-retrofitted homes.
We first define the ATT formally, then present the DiD design and its assumptions, and finally introduce the estimator used in this study, including covariate adjustments and matching procedures for constructing a comparable control group.

\subsubsection{The average treatment effect on the treated (ATT)}

The quantity we want to estimate is how much a retrofit reduces energy consumption.
In econometrics, this quantity is called the average treatment effect on the treated (ATT)~\cite{imbensNonparametricEstimationAverage2004}.
Let $t=0$ be the time before treatment and $t=1$ be the time after treatment. Let $Y_{i,t}(0)$ be the log of energy consumption for home $i$ at time $t$ if no treatment is done and $Y_{i,t}(1)$ be the log of energy consumption for home $i$ at time $t$ if a treatment is done.
Let $D_i$ be the indicator of home $i$ being in the treatment group. Then the ATT for home $i$ is defined as follows:

\begin{equation}
    \ATT =   \EE \left[Y_{i,1}(1) - Y_{i,1}(0) | D_i = 1\right].
\end{equation}

The first value, $Y_{i,1}(1)$, is the consumption we observe for treated homes.
The second value, $Y_{i,1}(0)$, is the counterfactual consumption, which we would observe for treated homes if the home had no retrofit done.

\subsubsection{Estimating the ATT with a DiD design}

Difference in differences (DiD)~\cite{ashenfelterUsingLongitudinalStructure1984} is a design used to estimate the causal effect of a treatment (here the retrofit) by comparing the changes in outcomes between different groups before and after the treatment. I
n its basic form, DiD takes the form of a 2 by 2 table, where the time periods are simply before and after the treatment and the two groups are the treated and the control group.
Difference in differences can be used to estimate the ATT: the counterfactual consumption is estimated with the consumption of the control group.

The main assumption of DiD is called the parallel trends assumption.
It states that, without treatment, the average of the change between before and after the treatment is the same for the treated and the control group.
Mathematically, this is defined as follows:

\begin{equation}
    \EE \left[ Y_{i,1}(0) - Y_{i,0}(0) | D_i = 0 \right] = \EE \left[ Y_{i,1}(0) - Y_{i,0}(0) | D_i = 1 \right].
\end{equation}

When this assumption is met, the DiD estimator $\delta$ for the ATT is defined as:
\begin{equation}
    \hat{\delta} = (\bar{y}_{1, 1} - \bar{y}_{0, 1}) - (\bar{y}_{1, 0} - \bar{y}_{0, 0}),
\end{equation}
where $\bar{y}_{t, g}$ is the average of the log of energy consumption for group $g$ ($g$ = 0 for control, $g$ = 1 for treated) at time $t$. It is a difference of differences, hence the name. This also corresponds to the estimate of $\beta_3$ in the linear regression model $Y_{i,t,g} = \beta_0 + \beta_1 \indicator_{t=1} + \beta_2 \indicator_{g=1} + \beta_3 \indicator_{t=1}\indicator_{g=1} + \varepsilon_{i,t}$.

However, this estimator is not consistent if the parallel trends assumption is not met. A weaker assumption is to ask for a parallel trend but conditionally on some covariates $X$. Mathematically, this is defined as follows:
\begin{equation}
    \EE \left[ Y_{i,1}(0) - Y_{i,0}(0) | D_i = 0, X_i \right] = \EE \left[ Y_{i,1}(0) - Y_{i,0}(0) | D_i = 1, X_i \right],
\end{equation}
where $X_i$ are the covariates for home $i$. In this case, adding $X_i$ to the linear regression model
\begin{equation}
    \label{eq:twfe}
    Y_{i,t,g} = \beta_0 + \beta_1 \indicator_{t=1} + \beta_2 \indicator_{g=1} + \beta_3 \indicator_{t=1}\indicator_{g=1} + \beta_4 X_i + \varepsilon_{i,t}
\end{equation}
and using $\hat \beta_3$ (the Ordinary Least Squares estimator of $\beta_3$) as the DiD estimator is called the Two-Way Fixed Effects (TWFE) estimator for the ATT~\cite{wooldridgeEconometricAnalysisCross2010}.

\subsubsection{Estimating the ATT with Doubly-Robust DiD}
Despite the presence of covariates, using a linear equation \eqref{eq:twfe} is not sufficient to remove most biases as described in \citep{caetanoDifferenceinDifferencesWhenParallel2024}.
We thus choose to use the state-of-the-art estimator called the \emph{doubly-robust estimator}, introduced in \cite*{santannaDoublyRobustDifferenceindifferences2020} and implemented as the R package `DRDID'.

Using the log-consumption as treatment outcome allows us to interpret the ATT as a percentage change in consumption.
The percent change in consumption is then $\left(e^{\text{ATT}} - 1\right)\times 100$.
This same transformation can be applied to the ATT, standard error, and lower and upper bounds of the confidence interval.

\paragraph{Choice of covariates}

The covariates used for ATT estimation in homes where the main heating is electrical are:
\begin{dashlist}[noitemsep, topsep=3pt]
    \item home surface (continuous)
    \item house age range (categorical)
    \item heating (categorical, electrical or heat pump)
    \item water heating (binary)
    \item swimming pool (binary)
    \item electric vehicle (binary)
    \item number of inhabitants (categorical)
    \item number of floors (categorical)
\end{dashlist}

For homes where the main heating is gas-powered, the only covariate is the home surface.
We made this choice due to the very small number of homes in this category, and adding more (categorical) variables made the estimation unstable.
The same choice was made for homes for the air-to-water heat pumps analysis.

This doubly robust estimator requires data structured around two time periods (pre- and post-treatment) and two groups (treated and control).
In the following section, we describe how we reduce the dataset to two time periods and define the control group.

\subsection{Reducing the time dimension: before and after the retrofit}

After correcting consumption data for temperature variations (see Section \ref{sec:methodology:weather_correction}), we construct the two time periods required for the DiD analysis: before and after the retrofit.

The procedure involves two aggregation steps:

\begin{enumerate}
    \item Monthly averaging: For each period (before/after retrofit), we compute the average consumption for each calendar month (January–December), pooling across years.
          This ensures that all months contribute equally to the definition of each time period.
    \item Household averaging: For each home, we then average the monthly consumption values separately for the pre- and post-retrofit periods.
          This ensures that each household contributes equally to the construction of the final time periods.
\end{enumerate}

Formally, for a given household, this can be expressed as:

\begin{equation}
    C_{t} = \frac{1}{12} \sum_{m=1}^{12} \left( \frac{1}{|N_{m,t}|}\sum_{y\in N_{m,t}} C_{m,y}^{\text{corrected}}\right)
\end{equation}
where $t =0,1$ is the time period (before/after retrofit), $N_{m,t}$ is the set of years with data in month $m$ and before retrofit ($N_{m,0}$) or after retrofit ($N_{m,1}$) and $C_{m,y}^{\text{corrected}}$ is the corrected consumption for year $y$ and month $m$ defined in Section \ref{sec:methodology:heating_degree_days_correction}

\subsection{Creation of the control group}
\label{sec:methodology:control_group_creation}

The estimator presented in the previous section relies on the existence of a control group.
As discussed, any group of homes with no renovations could in theory be used, but in order to further reduce bias, we present in this section how the control group was selected to be as close as possible to the treated group.

The pool of potential control homes is the set of homes from Hello Watt app users that did not declare any retrofit works.
For every home in the treated group, we select the $k$ most similar homes from the control pool. This is sometimes referred to as $k{:}1$ nearest neighbor matching.
The similarity is computed using a score that depends on self-reported home characteristics as presented in section \ref{sec:distance}.
We do not have a history of metadata for both groups, so we assume that they do not change over time.
For the specific case of heat pump homes, as it induces a change of metadata for the treated group, we enforce the control group to have electrical heating, rather than heat pump, as we want the control group to be as close as possible to the treated group before the retrofit.

In order to be matched, a control home must have consumption data available at least one year before and after the retrofit work of the home in the treated group, a condition similar to the one used in the selection of the homes in the treated group.
We must also have detected heating for the control home, both before and after the retrofit work, as it is demanded for the treated homes, cf. Section \ref{sec:data_selection}.

\subsubsection{Distance of similarity for homes with electric heating}
\label{sec:distance}

When studying electric consumption, the control group must respect some hard constraints on the characteristics: the heating system and the presence of a swimming pool must exactly match the one of the home in the treated group.

For the other characteristics, we define a distance as the sum of:
\begin{dashlist}[noitemsep, topsep=3pt]
    \item location: (distance in kilometers)/100 between the INSEE areas of homes, using ellipsoid model WGS-84
    \item inhabitants: absolute difference
    \item habitable surface: $\floor{10 \cdot \frac{\abs{s_1 - s_2}}{s_1}}$
    \item water heating: 0 if identical, 3 if different
    \item air conditioning: 0 if identical, 2 if different
    \item electric vehicle: 0 if identical, 4 if different
    \item home age: 0 if in the same group, 2 if different
    \item secondary heating (present or absent): 0 if identical, 3 if different
\end{dashlist}

\subsubsection{Distance of similarity for homes with gas heating}
As the gas consumption only consists of heating, water heating, and/or cooking, it is less affected by other variables.
Therefore, the distance of similarity for gas homes is similar but simpler than the one for electrical homes.

We implemented the same hard constraints for the control homes that the heating system must be gas, that they did not declare a retrofit, and that they have at least one year of gas consumption data before and after the retrofit of the corresponding home in the treated group.

Second, we build the following distance, again decomposed into smaller distances:
\begin{dashlist}[noitemsep, topsep=3pt]
    \item location: (distance in kilometers)/100 between the INSEE areas of homes, using ellipsoid model WGS-84.
    \item inhabitants: absolute difference
    \item habitable surface: $\floor{10 \cdot \frac{\abs{s_1 - s_2}}{s_1}}$
    \item water heating: 0 if identical, 3 if different
    \item home age: 0 if in the same group, 2 if different
    \item secondary heating (present or absent): 0 if identical, 1 if different
    \item swimming pool (present or absent): 0 if identical, 2 if different
\end{dashlist}

\subsubsection{Matching}

For each home in the treated group, we compute the distance of similarity with all homes in the control pool.
We then select the $k$ homes in the control pool that have the smallest distance of similarity with the given treated home.
In some rare cases, there are no homes in the control pool that match all the hard constraints, resulting in a control group that is slightly smaller than $k$ times the number of homes in the treated group.
The matching is done with replacement, meaning that one home in the control group can be selected for several treated homes.

\subsubsection{Construction of the before/after periods for the control group}

When a control home has a longer consumption history than its matched treated home, its time series is truncated to align with the observation window of the treated counterpart.
Each matched control home is then assigned the same (fictive) retrofit date as the treated home.
This ensures that the two periods (before and after retrofit) are defined relative to a common treatment date, thereby minimizing potential bias arising from heterogeneity in treatment timing across treated homes.

If a single control home is matched to multiple treated homes, distinct before/after periods are constructed for each match, since the reference treatment dates differ.
Consequently, the same control home may contribute multiple, non-identical before/after consumption profiles.

\section{Results}
\label{sec:results}

The objective of our analysis is to estimate the \emph{Average Treatment Effect on the Treated (ATT)}, defined as the expected reduction in energy consumption for households that undertook a retrofit, relative to what their consumption would have been in the absence of treatment.

We estimate separately the ATT by main heating energy source (gas or electricity) and type of retrofit measure (e.g.\ attic insulation, heat pump). Unless otherwise noted, we use $k=5$ nearest neighbors when constructing the control group, i.e. for each treated household, the five closest homes in the control group (based on observed characteristics) are selected.

Last, we perform an analysis of homes that switched from gas heating to electrical heat pumps, by analyzing at the same time the increase in electricity consumption before and after the switch, and the decrease in gas consumption.

\subsection{Results for electricity consumption}

A total of 1996 homes were selected in the treated group and 8851 homes in the control group.
The heat pumps considered in this analysis are restricted to air-to-air heat pumps; air-to-water heat pumps are not considered here, and are the focus of Section \ref{sec:heat_pump}.

Sample sizes for each retrofit measure are given in \Cref{tab:elec_sample_size}.

\begin{table}[h]
    \centering
    \footnotesize
    \begin{tabular}{lrr}
        \toprule
        retrofit measure    & treated & control \\
        \midrule
        changing windows    & 474     & 2087    \\
        comprehensive       & 113     & 515     \\
        heat pump           & 227     & 989     \\
        insulating attic    & 836     & 3722    \\
        insulating basement & 73      & 340     \\
        insulating walls    & 273     & 1198    \\
        \midrule
        total               & 1996    & 8851    \\
        \bottomrule
    \end{tabular}
    \caption{Number of homes in the treated and control groups for electricity consumption, for each retrofit measure.}
    \label{tab:elec_sample_size}
\end{table}

The comparison of consumptions between the two groups, before the retrofit, is given in \Cref{tab:consumptions_before_retrofits_elec}.

\subsubsection{Effect on the total consumption}

The results for the effect of retrofitting on electricity consumption when heating with electricity are presented in \Cref{fig:elec_results}. The table with the exact ATT estimates is given in \Cref{tab:elec_results} in the appendix.

\begin{figure}[h]
    \centering
    \includegraphics[width=0.48\textwidth]{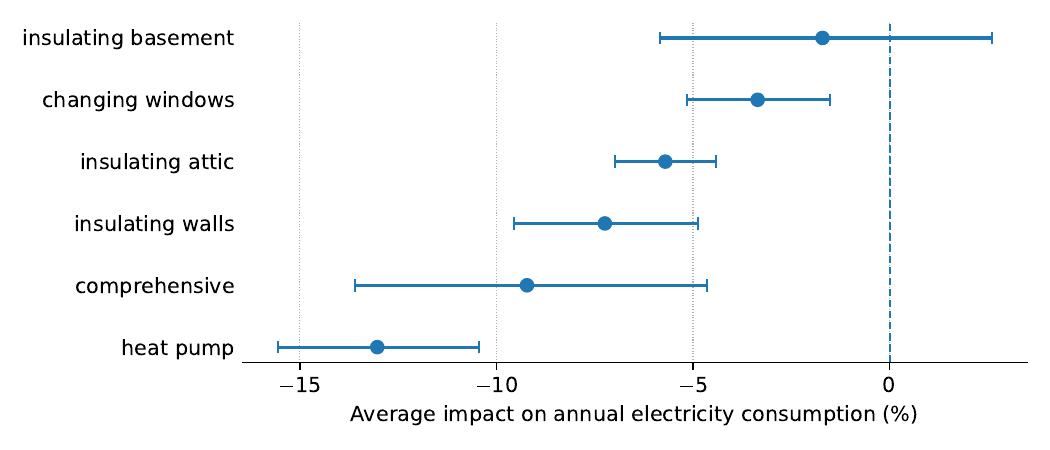}
    \caption{Average treatment effect on the treated (ATT) for electricity consumption in \%. It corresponds to the impact of several retrofit measures on the annual electricity consumption, for homes with electric heating. Error bars show the 95\% confidence interval.}
    \label{fig:elec_results}
\end{figure}

Except for basement insulation, which is not significant (ATT = -1.6\%), all other retrofit measures have a statistically significant effect on electricity consumption.
Changing the windows has the smallest significant effect (ATT = -3.3\%), while the comprehensive retrofit (i.e. two or more measures) has the second largest effect (ATT = -9.2\%).
The other two insulation measures also show significant reductions: attic insulation (ATT = -5.7\%), and wall insulation (ATT = -7.2\%).

Despite the fact that air-to-air heat pumps can be used as a cooling system in summer, they have a larger effect on electricity consumption (ATT = -13\%), the reduction in heating consumption being larger than the increase in consumption due to the air conditioning.

These values are close to those found in \cite{babamoussaEffetsLisolationThermique}, where the authors found an effect of -7.41\% for wall insulations, -5.36\% for attic insulations, -3.3\% for changing windows\footnote{The authors of \cite{babamoussaEffetsLisolationThermique} group measures in a slightly different way.
    They use the categories attic/roof/floor insulation, wall insulation, windows and French windows insulation, and comprehensive retrofit.}.
The measure that differs the most is the comprehensive retrofit, with a 4.65\% reduction on average, compared to -9.6\% for the comprehensive retrofit in this study.
However, their standard errors are large for this measure (3.5 percentage points). It is also surprising that the comprehensive retrofit has a smaller effect than the attic or wall insulation, the opposite of what we would expect.

To estimate the absolute effects of retrofitting on total consumption (expressed in kWh/year), we use the same approach as above but using the total consumption as the outcome variable instead of its logarithm. The results are given in \Cref{tab:elec_results_kwh}. Results are in line with relative effects.

The results range from -183 kWh/year for basement insulation (non-significant) to -1235 kWh/year for air-to-air heat pumps.
Using 79 g \COtwoeq /kWh \cite{Article10Arrete} for French electricity, the reduction in emissions for installing a heat pump corresponds to a reduction of 98 kg \COtwoeq /year on average.

\begin{table}[!ht]
    \centering
    \footnotesize
    \begin{tabular}{lrrrrr}
        \toprule
        retrofit measure    & ATT   & s.e. & LCB   & UCB  & p-value \\
        \midrule
        changing windows    & -253  & 103  & -455  & -52  & 0.014   \\
        comprehensive       & -938  & 283  & -1494 & -383 & <0.001  \\
        heat pump           & -1235 & 156  & -1540 & -929 & <0.001  \\
        insulating attic    & -563  & 74   & -707  & -418 & <0.001  \\
        insulating basement & -183  & 254  & -680  & 314  & 0.470   \\
        insulating walls    & -588  & 136  & -855  & -321 & <0.001  \\
        \bottomrule
    \end{tabular}
    \caption{Average treatment effect on the treated (ATT) for electricity consumption, in kWh/year. s.e. is the standard error, LCB is the lower bound of the 95\% confidence interval, and UCB is the upper bound.}
    \label{tab:elec_results_kwh}
\end{table}

\subsubsection{Effect on the heating consumption}
\label{sec:elec_results_heating}
For homes with electric heating, the part of electricity consumption that is due to heating is around 50\% of the total consumption.
Therefore, even if a retrofit measure has a large effect on heating consumption (in \%), the effect on the total consumption, expressed in \%, may be smaller.
This also makes it easier to compare the effect between electricity and gas, as on the other side, the gas consumption is mostly due to heating.

The decomposition between heating and non-heating (base) consumption is based on the estimator defined by \eqref{eq:heating_consumption} and \eqref{eq:base_consumption}.

\begin{figure}[!h]
    \centering
    \includegraphics[width=0.48\textwidth]{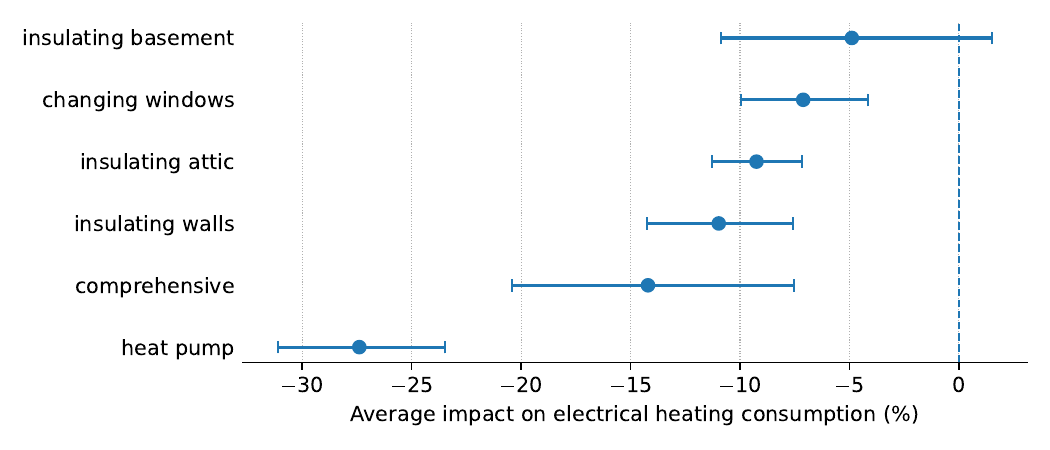}
    \caption{Average treatment effect on the treated (ATT) on the heating part of electricity consumption in \%. It corresponds to the impact of several retrofit measures on the annual electricity consumption, for homes with electric heating. Error bars show the 95\% confidence interval.}
    \label{fig:elec_results_heating}
\end{figure}

The results for the effect of retrofitting on heating consumption when heating with electricity are presented in \Cref{fig:elec_results_heating}.
As expected, the effect on heating only is larger than the effect on total consumption, with an effect up to -27 \% for heat pumps. Other retrofit measures also show a significant reduction in heating consumption, with an effect up to twice as large as the effect on total consumption.
The table with the exact ATT estimates is given in \Cref{tab:elec_results_heating} in the appendix.

For completeness, the results on the base part of the consumption are given in \Cref{tab:elec_results_base}.
The attic insulation retrofit is the only one with a significant effect on the base consumption, with an ATT of -3.9\%. All other retrofit measures have a non-significant effect on the base consumption.

\subsection{Results for gas consumption}

For gas-heated homes, the sample size is smaller than for electrical homes, with 423 homes in the treated group and 2,093 homes in the control group. The detailed sample sizes are given in \Cref{tab:gas_sample_size}.

This distribution is rather similar to the one for electrical homes, with the attic insulation having the largest sample size and the basement insulation the smallest.

\begin{table}[!ht]
    \centering
    \footnotesize
    \begin{tabular}{lrr}
        \toprule
        retrofit measure    & treated & control \\
        \midrule
        changing windows    & 110     & 539     \\
        comprehensive       & 32      & 159     \\
        insulating attic    & 187     & 927     \\
        insulating basement & 15      & 75      \\
        insulating walls    & 79      & 393     \\
        \midrule
        total               & 423     & 2093    \\
        \bottomrule
    \end{tabular}

    \caption{Number of homes in the treated and control groups for gas consumption, for each retrofit measure.}
    \label{tab:gas_sample_size}
\end{table}

The comparison of consumptions between the two groups, before the retrofit, is given in \Cref{tab:consumptions_before_retrofits_gas}.

\subsubsection{Effect on the total consumption}

The results for the effect of retrofit on gas consumption when heating with gas are presented in \Cref{fig:gas_results}. The exact ATT estimates are given in the appendix, in \Cref{tab:gas_results}.

\begin{figure}[h]
    \centering
    \includegraphics[width=0.48\textwidth]{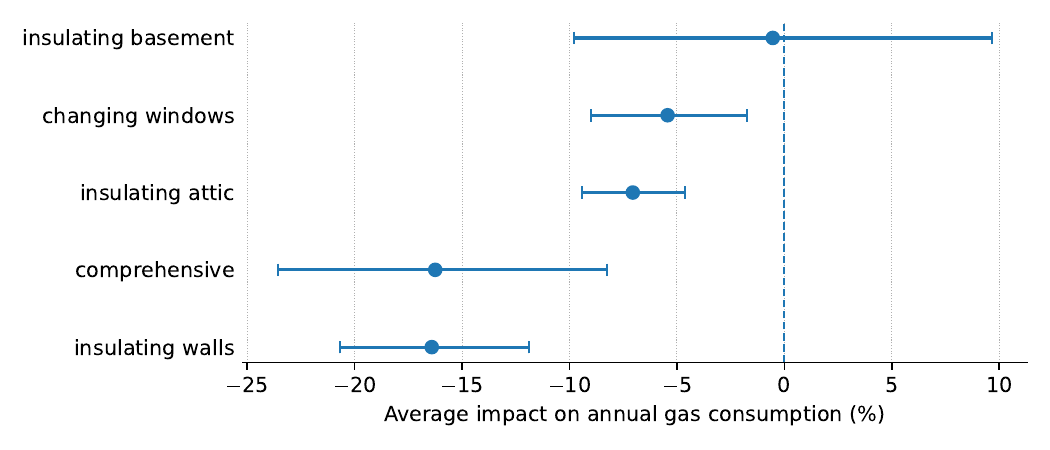}
    \caption{Average treatment effect on the treated (ATT) for gas consumption, in \%. It corresponds to the impact of several retrofit measures on the annual gas consumption, for homes with gas heating. Error bars show the 95\% confidence interval.}
    \label{fig:gas_results}
\end{figure}

Basement insulation shows a small, non-significant effect (ATT = -0.5\%), a result to interpret cautiously given the small treated sample (15 homes).
As for the electrical case, all other retrofit measures yield significant reductions in gas consumption.
Changing windows has a statistically significant reduction (ATT = -5.4\%).
Attic insulation and wall insulation both yield significant reductions (ATT = -7.05\% and -16.4\% respectively).
Comprehensive retrofit has an estimated effect of -16.2\%, with a large standard error (6.7 percentage points).

The authors of \cite{babamoussaEffetsLisolationThermique} found similar results: -13.86 \% for wall insulation, -8.27 \% for attic insulation and 5.9\% for changing windows.
Their estimate for comprehensive retrofit is larger than ours (-21.10\%).

We also computed for gas consumption the ATT in kWh/year, given in \Cref{tab:gas_results_kwh}. The results are in line with those in percents. The ATT, for positive significant effects, goes from -774 kWh/year for attic insulation to -1851 kWh/year for wall insulation.
Using a coefficient of 227 g \COtwoeq /kWh LHV \cite{Article10Arrete} for French gas and a 0.9 conversion factor from HHV to LHV, the reduction in emissions for comprehensive retrofit corresponds to a reduction from 158 kg \COtwoeq /year for the smallest significant effect (attic insulation) to 378 kg \COtwoeq /year for the largest significant effect (wall insulation).

\begin{table}[!ht]
    \centering
    \footnotesize
    \begin{tabular}{lrrrrr}
        \toprule
        retrofit measure    & ATT   & s.e. & LCB   & UCB   & p-value \\
        \midrule
        changing windows    & -854  & 302  & -1446 & -263  & 0.005   \\
        comprehensive       & -1858 & 583  & -3001 & -715  & 0.001   \\
        insulating attic    & -774  & 182  & -1131 & -417  & <0.001  \\
        insulating basement & 380   & 819  & -1226 & 1986  & 0.643   \\
        insulating walls    & -1851 & 307  & -2453 & -1249 & <0.001  \\
        \bottomrule
    \end{tabular}

    \caption{Average treatment effect on the treated (ATT) for gas consumption, in kWh HHV/year. s.e. is the standard error, LCB is the lower bound of the 95\% confidence interval, and UCB is the upper bound.}
    \label{tab:gas_results_kwh}
\end{table}

\subsubsection{Effect on the heating consumption}

\begin{figure}[!h]
    \centering
    \includegraphics[width=0.48\textwidth]{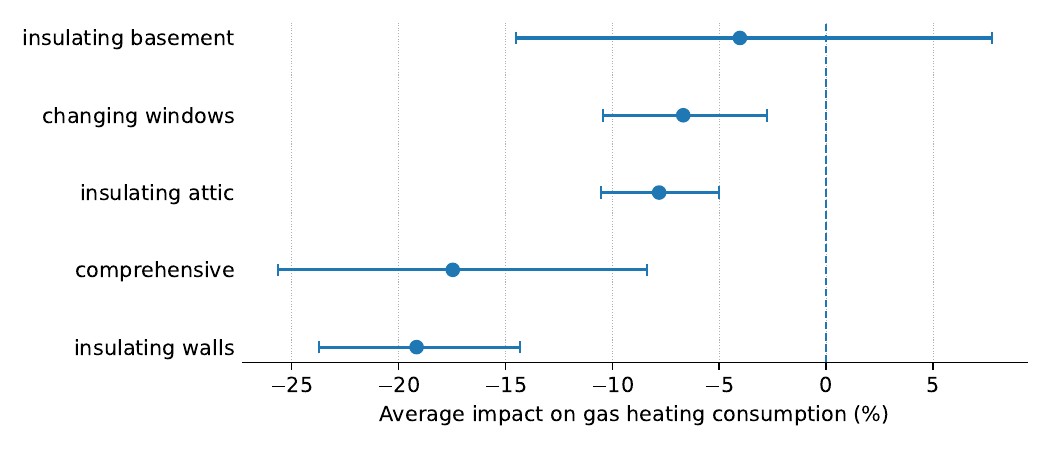}
    \caption{Average treatment effect on the treated (ATT) on the heating part of gas consumption in \%. It corresponds to the impact of several retrofit measures on the annual gas consumption, for homes with gas heating. Error bars show the 95\% confidence interval.}
    \label{fig:gas_results_heating}
\end{figure}

Similar to Section \ref{sec:elec_results_heating}, we computed the ATT for the heating consumption, presented in \Cref{fig:gas_results_heating}. In the case of gas consumption, heating represent around 80 \% of the total consumption, making the ATT for the heating consumption closer to the ATT for the total consumption. The exact values are given in \Cref{tab:gas_results_heating}.

For completeness, the results on the base part of the consumption are given in \Cref{tab:gas_results_base}, none being significant. As the base consumption is rather small, the change expressed in \% can be rather large without being significant.

When comparing the ATT on heating consumption between gas and electric heating, the estimates are rather similar, in particular for changing windows (-6.8\% for gas and -7.1\% for electric heating) and attic insulation (-7.8\% for gas and -9.2\% for electric heating). The ATT for insulating walls is larger in the gas case (19.6 \% v. 11.0 \%) but the confidence intervals overlap. Insulating basement is non-significant in both cases.

\subsection{Effects of replacing gas heating with an air-to-water heat pump}
We study in this section the effect of replacing a gas heating system (used for both heating and hot water) with an air-to-water heat pump.
\label{sec:heat_pump}
\subsubsection{Effect on the total consumption}

The results in this section are different from those of the two previous sections in that we do not limit the analysis to a single energy source, but rather consider both gas and electric heating at the same time to analyze the effect of replacing one with the other.

We thus consider homes for which we have electricity consumption data both before and after the retrofit, as well as gas consumption data before.
Gas consumption can be lacking after the retrofit for some homes, for instance if the gas meter had been closed.
In these cases, we assume the gas consumption to be zero.

A rather important share of homes is equipped with solar panels, which may have affected their electricity consumption and may have mitigated the increase in electricity consumption after the retrofit. We therefore decided to exclude these homes from the analysis to avoid biasing our results.

We restrict the sample to homes that match several conditions on their heating: we must detect gas heating before the retrofit, electric heating after the retrofit, and no gas heating after the retrofit.
Having some gas consumption after the retrofit does not result in exclusion, since gas is also used for cooking, but this consumption must not be thermosensitive.
This provides us with a group of homes that went from using only gas heating to using only electric heating.

Because of these different restrictions, the sample size is smaller than the one considered in the previous sections. We get 23 homes once the conditions are applied. We keep the same 5:1 ratio between treated and control groups, resulting in a total of 108 homes in the control group.

\begin{figure}[ht]
    \centering
    \includegraphics[width=0.48\textwidth]{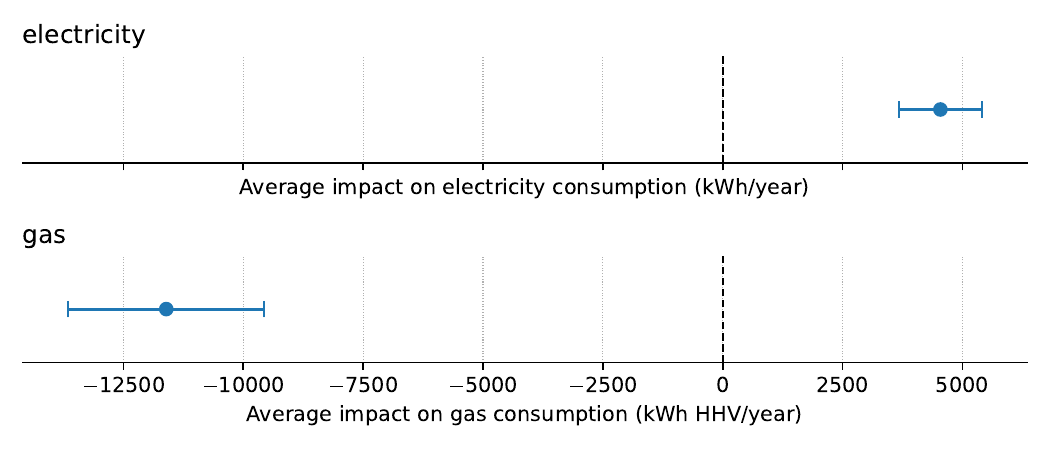}
    \caption{Average treatment effect on the treated (ATT), corresponding to the impact of switching from gas heating to heat pump, for electricity and gas consumption. Results in kWh HHV/year. The sample size is 24 homes for the treatment group, and 120 homes for the control group. Error bars show the 95\% confidence interval.}
    \label{fig:elec_gas_results_heat_pump}
\end{figure}

The results are presented in \Cref{fig:elec_gas_results_heat_pump}, which contains the ATT for both energy sources. The exact estimates are given in the appendix, in \Cref{tab:elec_gas_results_kwh}. The ATT is computed only in kWh/year (using the consumption as the outcome variable), as it is more relevant than the ATT in \% because of gas consumption becoming zero after the retrofit.  As heat pump is also used for domestic hot water, the ATT does not capture the pure effect of heating switching from gas to electricity, but both switching the heating and the hot water system.

The consumptions before the retrofit are given in \Cref{tab:consumptions_before_retrofits_heat_pump}.

\subsubsection{Estimation of \COtwo reduction}
We estimate the \COtwo reduction resulting from the retrofit by multiplying the ATT by the \COtwo emission factor of each energy source.
We use the French \COtwo emission factors from \cite{Article10Arrete}: 0.079 kg \COtwoeq/kWh for electricity (heating) and 0.227 kg \COtwoeq/kWh LHV for gas, and a factor of 0.9 to convert the gas consumption from HHV to LHV.

As detailed in \Cref{tab:co2_results_kgeqco2}, we observe a large reduction in \COtwo emissions of 2.0 tons \COtwoeq/year, which corresponds to a 85\% decrease.

As a comparison, the estimated \COtwo impact (full life cycle) of a air-to-water heat pump is around 8 tons \COtwoeq\footnote{The estimate is from \url{https://base-inies.fr/consultation/infos-produit/45148}. The INIES is a French database considered as the reference for the French energy sector.} less than four years of our estimated \COtwo reduction.
However, this comparison is limited since the heat pumps are primarily replacing old or failing gas boilers, which would have needed replacements anyways.

This reduction can be attributed to the combined effect of two factors.
The first is due to the heat pump performance, which generates more heat with the same amount of energy (by directly comparing the energies, the heat pump only uses 39\% of the energy used by the gas heater).
The second effect is due to the \COtwo emission per kWh, the emissions for gas being almost thrice those for electricity.

\begin{table}[ht]
    \centering
    \footnotesize
    \begin{tabular}{lrr}
        \toprule
        energy      & diff in consumption & diff. in emission \\
        \midrule
        electricity & 4540                & 359               \\
        gas         & -11606              & -2371             \\
        \midrule
        total       &                     & -2012             \\
        \bottomrule
    \end{tabular}

    \caption{Estimation of \COtwo reduction resulting from the air-to-water heat pump retrofit. The difference in consumption is computed in kWh/year and is the ATT of the retrofit. A negative value means a reduction in consumption after the retrofit. The difference in emission is computed in kg \COtwoeq/year. }
    \label{tab:co2_results_kgeqco2}
\end{table}

\section{Discussion}
\label{sec:discussion}

\subsection{Limitations}
We identify several limits to our approaches, that we tried to mitigate as much as possible.

\begin{dashlist}
    \item Reducing the time periods to only two (before and after the retrofit) may introduce a bias as all homes are not treated at the same time.
    This makes itdifficult to estimate the (natural) trend of consumption, as it is not the same for all homes.
    This is due to a lack of consumption data, as each home has a different coverage period.

    We tried to mitigate this by making the control group as close as possible to the treated group, in particular by using the same time periods when computing the consumption before and after the retrofit.
    \item Another consequence of reducing the time period is that it eliminates the possibility of estimating the evolution of the retrofit effect over time after treatment.
    This approach is, for instance, used in \cite{babamoussaEffetsLisolationThermique}.
    We tried to use a synthetic time period (creating a synthetic time, relative to the treatment time) to estimate the trend of the ATT, but this leads to insignificant results.
    This is due to the fact that energy consumption is highly seasonal, and applying a different time shift to each home mixes the different seasonal patterns.
    \item  Sample size is small for some retrofit measures, especially for gas-heated homes, which makes it difficult to generalize the results.
    \item All homes (treated and control) are occupied by users of the Hello Watt app, which may introduce a selection bias as Hello Watt users may not be representative of the general French population.
    \item All data relative to the retrofit and the home characteristics are directly provided by Hello Watt users.
    Users can declare partial or incorrect information, which may bias the results, for instance, for the date of the retrofit.
    Since there is no way to positively declare that a user has not done a retrofit, the control group relies solely on the absence of a retrofit statement; some users in the control group may have undertaken a retrofit without our knowledge.
    \item Second homes, which can also be declared on the Hello Watt app, may bias results; such homes are often only occasionally occupied, and if the occupancy pattern of a home changes between pre- and post-treatment periods, observed differences in consumption may reflect changes in occupancy rather than the impact of the retrofit.
\end{dashlist}

\subsection{Discrepancy from Theoretical Estimates}
Our doubly-robust difference-in-differences analysis demonstrates that the decreases in both gas and electricity consumption following many retrofits are significant both statistically and in effect.
However, the consumption decreases we observed are lower than theoretical estimates \cite{RenovationEnergetiqueMaisons}.
This study, called Tremi 2020, was made by the French Ministry for Ecological Transition.
Using a theoretical approach (the DPE 3CL), it estimated the reduction in final energy for each type of retrofit: 0.5 MWh/year for changing windows, 1.6 MWh/year for basement insulation, 1.6 MWh/year for attic insulation, 2.9 MWh/year for wall insulation, and 13.5 MWh/year for heat pump installation.
The study does not differentiate between gas and electricity.
These values are larger than the effect we observe in our study, especially in the electrical case, where the effects can be up to 5.5 times smaller (for wall insulation).
Their estimates are, however, closer to the ATT we observe in the gas case, and in particular, our estimates for changing windows are larger than theirs (1 MWh/year).
Nevertheless, our results are consistent with the results of \cite{babamoussaEffetsLisolationThermique}, which also uses real consumption data and observes a similar discrepancy from the theoretical estimates.
Also, the reduction in electricity we observe after installation of an air-to-air heat pump (-13\% of the total consumption, -27\% of the heating consumption) is close to the raw results of \cite{dupretEtudeConsommationsPAC2025} (-18\% of the total consumption, -34\% of the heating consumption) before they correct for vacancy, climate, and extra heating. Although we correct for climate, we don't take into account vacancies and supplementary heating sources, which is part of the limitations of this study we discussed in the previous section, and may explain in part why the estimate is lower than expected.

The theoretical estimates of retrofits are based on several hypotheses, the main one being that all other factors are held constant, especially user behavior, but there is a range of user variables which could lower the empirical estimates.
Another hypothesis concerns the geometry of the home, using only the surface area to estimate it.
One of the main suspects for the discrepancy is the rebound effect, in which the \textit{behavior} of the occupants changes in response to the retrofit \cite{fontvivancoReboundEffectSustainability2022,sorrellLimitsEnergySufficiency2020}.
For example, a home might increase its heating consumption after a retrofit because its upgraded insulation makes heating more affordable.
This is called the \textit{prebound effect}, when the consumption prior to the retrofit is lower than expected consumption, because of the price of energy \cite{galvinQuantificationPreboundEffects2016}.
This prebound effect also relates to the fact that in France, homes with low-rated energy performance certificates (\textit{Diagnostic de performance énergétique}, DPE) have a lower consumption than expected \cite{matheronComparerDPEConsommation2023}.
Under this hypothesis, the benefits of retrofits are distributed, for example, between energy consumption and home comfort, instead of being concentrated in just one metric.
There are also confounding variables, such as regulations and the price of energy, which could also affect user behavior across the time span necessary for a two-by-two methodology.

The cause is possibly a combination of several factors.
At the moment, we lack definitive studies in France on the rebound effect and user behavior to make any definitive assertions about the balance of variables contributing to the discrepancy.
Other actors, such as ADEME, with, for instance, the study Panel Elecdom, are running new studies in this direction \cite{PanelUsagesElectrodomestiques}.

Hello Watt's current projects may allow us to extend our research in this direction as well.
These initiatives include
\begin{enumerate}
    \item \textit{Home thermostat access}.
          Hello Watt will soon be installing and have access to home thermostat data as part of its goal of being the all-in-one app for controlling one's home.
          This data will help to explicitly measure the rebound effect after retrofits.

    \item \textit{Hello Watt performing more retrofits for users}.
          As mentioned above, dates and types of retrofits can be unreliable because they are entirely provided by the user.
          Hello Watt has more precise data and metadata about the home retrofits which Hello Watt itself performs; at the moment we do not have enough works with the requisite one year of post-retrofit consumption data, but once we do we will be able to combine consumption data with precise metadata to increase the power of our study.
          Furthermore, the app is continuously acquiring more users who have performed or are performing renovations with other actors (accelerated, for example, through Hello Watt's collaborations with partners such as Société Générale and Crédit Agricole), the larger sample size will enable more precise studies in the future, especially for renovations for which most actors do not have much data.
\end{enumerate}

\section{Conclusion}

Our findings show that single retrofit measures—such as attic and wall insulation, as well as window replacements—lead to statistically significant reductions in household energy consumption, especially for homes with gas heating.
However, these realized savings are systematically lower than theoretical estimates, likely due to behavioral factors and other confounding influences discussed above.
The effect sizes are in line with previous empirical studies and confirm that retrofitting remains an effective tool for energy demand reduction at the population level, even if not all expected savings are captured in practice.

Heat pump installations are unique among the retrofits we consider. Air-to-air heat pumps result in a 13\% reduction in overall electricity consumption, even when taking into account the increase in electricity consumption due to the air conditioning.
Concerning air-to-water heat pumps, as discussed in Section \ref{sec:heat_pump}, decarbonization is one of the dimensions in which the gains from heat pumps are distributed.
Even if the gains purely in terms of efficiency are not at their theoretical maximum, transitioning away from gas significantly reduces \COtwo emissions, with an estimate of 87\% of \COtwoeq reduction.
Current geopolitical tensions are also accelerating the demand for Europe's energy sovereignty\cite{REPowerEU2022}, and movement away from fossil fuels is a crucial part of this agenda.

Hello Watt provides a range of tools to its users to reduce their consumption and carbon footprint.
In addition to making retrofits cheaper and more accessible, the Hello Watt app helps users better understand their energy usage and take actions to reduce it.
Among others, the app provides: machine learning models which disaggregate consumption data into categories (such as heating, water heating, cooking) to identify potential savings \cite{culiereBayesianModelElectrical2020c,belikovDomainKnowledgeAids2022}; personalized alerts when consumption exceeds thresholds; a comparator which helps users understand their consumption relative to other homes with similar metadata; a convenient means to compare and change energy providers, in particular energy providers with a focus on clean energy. Hello Watt is also investigating the effect of following one's consumption regularly on the app on energy consumption, and preliminary results have found a statistically significant correlation; a study quantifying this effect is currently in progress.


\bibliographystyle{elsarticle-num}
\bibliography{renovation}
\pagebreak

\appendix
\appendix
\counterwithin{table}{section}
\renewcommand{\thetable}{\Alph{section}.\arabic{table}}  
\setcounter{table}{0}  

\section{Sample characteristics}
\label{sec:stats_homes}
We detail here the descriptive statistics of the samples. \Cref{tab:house_characteristics_elec} shows the descriptive statistics for the electrical case, while \Cref{tab:house_characteristics_gas} shows the descriptive statistics for the gas case.

\Cref{tab:house_characteristics_heat_pump} shows the descriptive statistics for the air-to-water heat pump case. As heating and water heating variables are not relevant due to the change in heating source, they are not shown in the table, contrary to the other two cases.

\begin{table}[!ht]
    \centering
    \footnotesize
    \begin{tabular}{llrr}
        \toprule
                                                 &                 & Control & Treated \\

        \midrule
        \multirow[t]{2}{*}{home type}            & flat            & 7.3     & 8.2     \\
                                                 & house           & 92.7    & 91.8    \\
        \cline{1-4}
        \multirow[t]{3}{*}{floor count}          & 1               & 31.8    & 33.1    \\
                                                 & 2               & 60.8    & 57.9    \\
                                                 & 3 or more       & 7.4     & 9.0     \\
        \cline{1-4}
        \multirow[t]{4}{*}{inhabitants}          & 1               & 9.4     & 12.1    \\
                                                 & 2               & 45.5    & 39.5    \\
                                                 & 3               & 16.7    & 17.2    \\
                                                 & 4               & 28.4    & 31.2    \\
        \cline{1-4}
        \multirow[t]{10}{*}{water heating}       & elec            & 85.8    & 81.2    \\
                                                 & gas             & 0.1     & 0.3     \\
                                                 & heat pump       & 3.8     & 4.5     \\
                                                 & missing         & 0.1     & 0.0     \\
                                                 & oil             & 0.1     & 0.2     \\
                                                 & other           & 1.5     & 2.3     \\
                                                 & propane         & 0.0     & 0.1     \\
                                                 & shared          & 0.0     & 0.2     \\
                                                 & solar           & 0.6     & 1.5     \\
                                                 & thermodynamic   & 7.8     & 9.9     \\
        \cline{1-4}
        \multirow[t]{7}{*}{house age range}      & 1919 and before & 8.3     & 10.2    \\
                                                 & 1919 1945       & 3.6     & 5.0     \\
                                                 & 1946 1970       & 9.2     & 11.8    \\
                                                 & 1971 1990       & 39.5    & 37.9    \\
                                                 & 1991 2005       & 19.3    & 19.1    \\
                                                 & 2006 now        & 19.4    & 15.1    \\
                                                 & other           & 0.7     & 0.8     \\
        \cline{1-4}
        \multirow[t]{2}{*}{heating}              & elec            & 67.9    & 67.4    \\
                                                 & heat pump       & 32.1    & 32.6    \\
        \cline{1-4}
        \multirow[t]{10}{*}{secondary heating}   & elec            & 3.9     & 4.2     \\
                                                 & fireplace       & 5.8     & 5.6     \\
                                                 & gas             & 0.0     & 0.1     \\
                                                 & heat pump       & 3.2     & 3.1     \\
                                                 & none/missing    & 76.8    & 75.1    \\
                                                 & oil             & 0.3     & 0.3     \\
                                                 & other           & 0.8     & 1.1     \\
                                                 & pellet stove    & 3.3     & 4.3     \\
                                                 & propane         & 0.0     & 0.1     \\
                                                 & wood stove      & 5.9     & 6.2     \\
        \cline{1-4}
        \multirow[t]{2}{*}{has swimming pool}    & False           & 84.3    & 81.9    \\
                                                 & True            & 15.7    & 18.1    \\
        \cline{1-4}
        \multirow[t]{2}{*}{has electric vehicle} & False           & 91.8    & 89.4    \\
                                                 & True            & 8.2     & 10.6    \\
        \cline{1-4}
        \multirow[t]{2}{*}{home surface}         & mean            & 125.0   & 124.3   \\
                                                 & std             & 42.2    & 47.2    \\
        \cline{1-4}
        sample size                              &                 & 8851    & 1996    \\
        \cline{1-4}
        \bottomrule
    \end{tabular}

    \caption{Home characteristics for homes with electric heating. All different retrofit are aggregated. Results are in \% for each variable, except for home surface which is in m$^2$.}
    \label{tab:house_characteristics_elec}
\end{table}

\begin{table}[!ht]
    \centering
    \footnotesize
    \begin{tabular}{llrr}
        \toprule
                                                 &                 & Control & Treated \\

        \midrule
        \multirow[t]{2}{*}{home type}            & flat            & 7.1     & 7.3     \\
                                                 & house           & 92.8    & 92.7    \\
        \cline{1-4}
        \multirow[t]{3}{*}{floor count}          & 1               & 14.8    & 16.1    \\
                                                 & 2               & 65.3    & 65.7    \\
                                                 & 3 or more       & 19.9    & 18.2    \\
        \cline{1-4}
        \multirow[t]{4}{*}{inhabitants}          & 1               & 7.8     & 10.6    \\
                                                 & 2               & 44.4    & 38.3    \\
                                                 & 3               & 17.4    & 18.9    \\
                                                 & 4               & 30.4    & 32.2    \\
        \cline{1-4}
        \multirow[t]{7}{*}{water heating}        & elec            & 11.1    & 11.1    \\
                                                 & gas             & 86.9    & 84.6    \\
                                                 & missing         & 0.1     & 0.0     \\
                                                 & other           & 0.0     & 0.5     \\
                                                 & propane         & 0.0     & 0.2     \\
                                                 & solar           & 0.8     & 1.7     \\
                                                 & thermodynamic   & 1.1     & 1.9     \\
        \cline{1-4}
        \multirow[t]{7}{*}{house age range}      & 1919 and before & 12.9    & 13.5    \\
                                                 & 1919 1945       & 11.8    & 12.1    \\
                                                 & 1946 1970       & 25.5    & 26.0    \\
                                                 & 1971 1990       & 26.7    & 26.7    \\
                                                 & 1991 2005       & 16.3    & 15.6    \\
                                                 & 2006 now        & 6.7     & 5.4     \\
                                                 & other           & 0.2     & 0.7     \\
        \cline{1-4}
        \multirow[t]{2}{*}{heating}              & gas             & 99.9    & 100.0   \\
                                                 & missing         & 0.1     & 0.0     \\
        \cline{1-4}
        \multirow[t]{9}{*}{secondary heating}    & elec            & 12.7    & 4.3     \\
                                                 & fireplace       & 6.5     & 3.3     \\
                                                 & gas             & 0.6     & 0.0     \\
                                                 & heat pump       & 5.4     & 2.1     \\
                                                 & none/missing    & 61.8    & 84.6    \\
                                                 & other           & 1.6     & 0.0     \\
                                                 & pellet stove    & 4.0     & 2.1     \\
                                                 & propane         & 0.0     & 0.0     \\
                                                 & wood stove      & 7.5     & 3.5     \\
        \cline{1-4}
        \multirow[t]{2}{*}{has swimming pool}    & False           & 85.3    & 84.9    \\
                                                 & True            & 14.7    & 15.1    \\
        \cline{1-4}
        \multirow[t]{2}{*}{has electric vehicle} & False           & 87.2    & 83.5    \\
                                                 & True            & 12.8    & 16.5    \\
        \cline{1-4}
        \multirow[t]{2}{*}{home surface}         & mean            & 131.6   & 130.7   \\
                                                 & std             & 45.8    & 46.9    \\
        \cline{1-4}
        sample size                              &                 & 2093    & 423     \\
        \cline{1-4}
        \bottomrule
    \end{tabular}
    \caption{House characteristics for homes with gas heating. All different retrofits are aggregated. Results are in \% for each variable, except for home surface which is in m$^2$.}
    \label{tab:house_characteristics_gas}
\end{table}

\begin{table}[htbp]
    \centering
    \footnotesize
    \begin{tabular}{llrr}
        \toprule
                                                 &                 & Control & Treated \\

        \midrule
        home type                                & house           & 100.0   & 100.0   \\
        \cline{1-4}
        \multirow[t]{3}{*}{floor count}          & 1               & 11.1    & 21.7    \\
                                                 & 2               & 73.1    & 65.2    \\
                                                 & 3 or more       & 15.7    & 13.0    \\
        \cline{1-4}
        \multirow[t]{5}{*}{inhabitants}          & 2               & 26.9    & 30.4    \\
                                                 & 3               & 22.2    & 13.0    \\
                                                 & 4               & 39.8    & 43.5    \\
                                                 & 5               & 9.3     & 13.0    \\
                                                 & 6               & 1.9     & 0.0     \\
        \cline{1-4}
        \multirow[t]{6}{*}{house age range}      & 1919 and before & 10.2    & 8.7     \\
                                                 & 1919 1945       & 15.7    & 13.0    \\
                                                 & 1946 1970       & 13.0    & 13.0    \\
                                                 & 1971 1990       & 19.4    & 21.7    \\
                                                 & 1991 2005       & 19.4    & 26.1    \\
                                                 & 2006 now        & 22.2    & 17.4    \\
        \cline{1-4}
        \multirow[t]{7}{*}{secondary heating}    & elec            & 12.0    & 0.0     \\
                                                 & fireplace       & 10.2    & 13.0    \\
                                                 & heat pump       & 8.3     & 0.0     \\
                                                 & none/missing    & 58.3    & 73.9    \\
                                                 & other           & 0.9     & 0.0     \\
                                                 & pellet stove    & 2.8     & 4.3     \\
                                                 & wood stove      & 7.4     & 8.7     \\
        \cline{1-4}
        \multirow[t]{2}{*}{has swimming pool}    & False           & 85.2    & 82.6    \\
                                                 & True            & 14.8    & 17.4    \\
        \cline{1-4}
        \multirow[t]{2}{*}{has electric vehicle} & False           & 83.3    & 60.9    \\
                                                 & True            & 16.7    & 39.1    \\
        \cline{1-4}
        \multirow[t]{2}{*}{home surface}         & mean            & 140.7   & 140.0   \\
                                                 & std             & 38.7    & 40.9    \\
        \cline{1-4}
        sample size                              &                 & 108.0   & 23.0    \\
        \cline{1-4}
        \bottomrule
    \end{tabular}
    \caption{House characteristics for homes for the air-to-water heat pump case. Results are in \% for each variable, except for home surface which is in m$^2$.}
    \label{tab:house_characteristics_heat_pump}
\end{table}

\pagebreak

\section{Consumption of the treated and control groups before retrofits}

\subsection{Electric heating}

We present in \Cref{tab:consumptions_before_retrofits_elec} the mean electricity consumption of the treated and control groups before retrofits, for each retrofit measure. The consumptions are very similar between the treated and control groups in this case

\begin{table}[!ht]
    \centering
    \footnotesize
    \begin{tabular}{lrr}
        \toprule
                            & Control & Treated \\
        \midrule
        changing windows    & 10689   & 10614   \\
        comprehensive       & 10386   & 10952   \\
        heat pump           & 10573   & 10589   \\
        insulating attic    & 10735   & 11104   \\
        insulating basement & 11342   & 11468   \\
        insulating walls    & 10321   & 10018   \\
        \bottomrule
    \end{tabular}

    \caption{Mean electricity consumption of the treated and control groups before retrofits for the electricity case, in kWh/year.}
    \label{tab:consumptions_before_retrofits_elec}
\end{table}

\subsection{Gas heating}
We present in \Cref{tab:consumptions_before_retrofits_gas} the mean gas consumption of the treated and control groups before retrofits, for each retrofit measure.
\begin{table}[!ht]
    \centering
    \footnotesize
    \begin{tabular}{lrr}
        \toprule
                            & Control & Treated \\
        \midrule
        changing windows    & 15183   & 16152   \\
        comprehensive       & 15098   & 16484   \\
        insulating attic    & 14941   & 15032   \\
        insulating basement & 19449   & 16014   \\
        insulating walls    & 14322   & 14230   \\
        \bottomrule
    \end{tabular}

    \caption{Mean gas consumption of the treated and control groups before retrofits for the gas case, in kWh HHV/year.}
    \label{tab:consumptions_before_retrofits_gas}
\end{table}

\subsection{Air-to-water heat pump}

We present in \Cref{tab:consumptions_before_retrofits_heat_pump} the mean gas consumption of the treated and control groups before retrofits, for the air-to-water heat pump retrofit. In this case, both groups have similar consumptions, for both energy sources.

\begin{table}[!ht]
    \centering
    \footnotesize
    \begin{tabular}{lrr}
        \toprule
                    & Control & Treated \\
        \midrule
        electricity & 5229    & 4927    \\
        gas         & 13754   & 13519   \\
        \bottomrule
    \end{tabular}
    \caption{Mean electricity and gas consumptions of the treated and control groups before retrofits for the air-to-water heat pump case, in kWh/year for electricity and in kWh HHV/year for gas.}
    \label{tab:consumptions_before_retrofits_heat_pump}
\end{table}

\pagebreak

\section{Exact ATT estimates}
\label{sec:appendix:exact_att}
This section presents the ATT estimates for each retrofit measure, using the DRDID estimator \cite{santannaDoublyRobustDifferenceindifferences2020}.

\subsection{Electric heating}
The ATT estimates for each retrofit measure, corresponding to \Cref{fig:elec_results} are given in \Cref{tab:elec_results}.
\begin{table}[!ht]
    \centering
    \footnotesize
    \begin{tabular}{lrrrrr}
        \toprule
        retrofit measure    & ATT    & s.e. & LCB    & UCB    & p-value \\
        \midrule
        changing windows    & -3.36  & 0.97 & -5.16  & -1.52  & <0.001  \\
        comprehensive       & -9.22  & 2.55 & -13.59 & -4.64  & <0.001  \\
        heat pump           & -13.03 & 1.51 & -15.55 & -10.43 & <0.001  \\
        insulating attic    & -5.70  & 0.70 & -6.98  & -4.41  & <0.001  \\
        insulating basement & -1.70  & 2.21 & -5.83  & 2.60   & 0.432   \\
        insulating walls    & -7.24  & 1.29 & -9.54  & -4.88  & <0.001  \\
        \bottomrule
    \end{tabular}

    \caption{Average treatment effect on the treated (ATT) for electricity consumption, in \%. s.e. is the standard error, LCB is the lower bound of the 95\% confidence interval, and UCB is the upper bound.}
    \label{tab:elec_results}
\end{table}

The total electricity consumption is split into heating and base consumption, with the heating consumption being around 50\% of the total consumption, using the estimator defined in Equations \ref{eq:heating_consumption} and \ref{eq:base_consumption}.

The ATT estimates for each retrofit measure corresponding to \Cref{fig:elec_results_heating}, on the heating consumption are given in \Cref{tab:elec_results_heating}.

\begin{table}[!ht]
    \centering
    \footnotesize
    \begin{tabular}{lrrrrr}
        \toprule
        retrofit measure    & ATT    & s.e. & LCB    & UCB    & p-value \\
        \midrule
        changing windows    & -7.11  & 1.60 & -9.97  & -4.17  & <0.001  \\
        comprehensive       & -14.20 & 3.90 & -20.40 & -7.53  & <0.001  \\
        heat pump           & -27.39 & 2.71 & -31.10 & -23.48 & <0.001  \\
        insulating attic    & -9.25  & 1.15 & -11.27 & -7.18  & <0.001  \\
        insulating basement & -4.89  & 3.38 & -10.88 & 1.50   & 0.131   \\
        insulating walls    & -10.97 & 1.93 & -14.24 & -7.58  & <0.001  \\
        \bottomrule
    \end{tabular}

    \caption{Average treatment effect on the treated (ATT) for the electrical heating consumption, in \%. s.e. is the standard error, LCB is the lower bound of the 95\% confidence interval, and UCB is the upper bound.}
    \label{tab:elec_results_heating}
\end{table}

Finally, the ATT estimates for each retrofit measure on the base consumption are given in \Cref{tab:elec_results_base}. As it can be expected, the ATT estimates are close to zero, with the attic insulation retrofit which has a significant effect.

\begin{table}[!ht]
    \centering
    \footnotesize
    \begin{tabular}{lrrrrr}
        \toprule
        retrofit measure    & ATT   & s.e. & LCB   & UCB   & p-value \\
        \midrule
        changing windows    & -0.11 & 1.24 & -2.50 & 2.34  & 0.927   \\
        comprehensive       & -4.31 & 3.09 & -9.85 & 1.57  & 0.147   \\
        heat pump           & -1.78 & 2.03 & -5.58 & 2.18  & 0.373   \\
        insulating attic    & -3.86 & 1.12 & -5.93 & -1.75 & <0.001  \\
        insulating basement & 0.32  & 3.66 & -6.51 & 7.65  & 0.929   \\
        insulating walls    & -3.31 & 1.99 & -6.97 & 0.49  & 0.087   \\
        \bottomrule
    \end{tabular}

    \caption{Average treatment effect on the treated (ATT) for the electrical base consumption, in \%. s.e. is the standard error, LCB is the lower bound of the 95\% confidence interval, and UCB is the upper bound.}
    \label{tab:elec_results_base}
\end{table}

\subsection{Gas heating}

The ATT estimates for each retrofit measure, using the DRDID estimator \cite{santannaDoublyRobustDifferenceindifferences2020}, corresponding to \Cref{fig:gas_results} are given in \Cref{tab:gas_results}.

\begin{table}[!ht]
    \centering
    \footnotesize

    \begin{tabular}{lrrrrr}
        \toprule
        retrofit measure    & ATT    & s.e. & LCB    & UCB    & p-value \\
        \midrule
        changing windows    & -5.44  & 1.97 & -8.98  & -1.76  & 0.004   \\
        comprehensive       & -16.25 & 4.77 & -23.56 & -8.24  & <0.001  \\
        insulating attic    & -7.05  & 1.32 & -9.42  & -4.63  & <0.001  \\
        insulating basement & -0.54  & 5.10 & -9.79  & 9.65   & 0.913   \\
        insulating walls    & -16.41 & 2.72 & -20.69 & -11.90 & <0.001  \\
        \bottomrule
    \end{tabular}

    \caption{Average treatment effect on the treated (ATT) for gas consumption, in \%. s.e. is the standard error, LCB is the lower bound of the 95\% confidence interval, and UCB is the upper bound.}
    \label{tab:gas_results}
\end{table}

The total gas consumption is split into heating and base consumption, with the heating consumption being around 75\% of the total consumption, using the estimator defined in Equations \ref{eq:heating_consumption} and \ref{eq:base_consumption}.

The ATT estimates for the heating part of the gas consumption are given in \Cref{tab:gas_results_heating}.

\begin{table}[!ht]
    \centering
    \footnotesize
    \begin{tabular}{lrrrrr}
        \toprule
        retrofit measure    & ATT    & s.e. & LCB    & UCB    & p-value \\
        \midrule
        changing windows    & -6.69  & 2.12 & -10.45 & -2.76  & <0.001  \\
        comprehensive       & -17.46 & 5.46 & -25.63 & -8.40  & <0.001  \\
        insulating attic    & -7.81  & 1.55 & -10.55 & -5.00  & <0.001  \\
        insulating basement & -4.04  & 6.08 & -14.53 & 7.74   & 0.485   \\
        insulating walls    & -19.15 & 3.02 & -23.73 & -14.30 & <0.001  \\
        \bottomrule
    \end{tabular}

    \caption{Average treatment effect on the treated (ATT) for the gas heating consumption, in \%. s.e. is the standard error, LCB is the lower bound of the 95\% confidence interval, and UCB is the upper bound.}
    \label{tab:gas_results_heating}
\end{table}

Finally, the ATT estimates for each retrofit measure on the base consumption are given in \Cref{tab:gas_results_base}. As it can be expected, the ATT estimates are close to zero, with the attic insulation retrofit which has a significant effect.

\begin{table}[!ht]
    \centering
    \footnotesize
    \begin{tabular}{lrrrrr}
        \toprule
        retrofit measure    & ATT    & s.e.  & LCB    & UCB   & p-value \\
        \midrule
        changing windows    & 13.42  & 15.00 & -13.76 & 49.18 & 0.368   \\
        comprehensive       & -28.02 & 24.10 & -52.86 & 9.91  & 0.128   \\
        insulating attic    & -5.89  & 6.07  & -16.16 & 5.64  & 0.303   \\
        insulating basement & 12.43  & 15.48 & -15.21 & 49.07 & 0.416   \\
        insulating walls    & -14.26 & 8.32  & -26.70 & 0.29  & 0.054   \\
        \bottomrule
    \end{tabular}

    \caption{Average treatment effect on the treated (ATT) for the gas base consumption, in \%. s.e. is the standard error, LCB is the lower bound of the 95\% confidence interval, and UCB is the upper bound.}
    \label{tab:gas_results_base}
\end{table}

\subsection{Air-to-water heat pump}
The \Cref{tab:elec_gas_results_kwh} presents the ATT estimates for electricity and gas consumption, for switching from gas heating to heat pump, using the DRDID method, corresponding to \Cref{fig:elec_gas_results_heat_pump}.
\begin{table}[!ht]
    \centering
    \footnotesize
    \begin{tabular}{lrrrrr}
        \toprule
        energy      & ATT    & s.e. & LCB    & UCB   & p-value \\
        \midrule
        electricity & 4447   & 439  & 3586   & 5309  & <0.001  \\
        gas         & -11468 & 1025 & -13477 & -9459 & <0.001  \\
        \bottomrule
    \end{tabular}

    \caption{Average treatment effect on the treated (ATT) for electricity and gas consumption, for switching from gas heating to heat pump. Results in kWh/year. The sample size is 24 homes for the treatment group, and 120 homes for the control group. s.e. is the standard error, LCB is the lower bound of the 95\% confidence interval, and UCB is the upper bound.}
    \label{tab:elec_gas_results_kwh}
\end{table}

\pagebreak

\section{TWFE estimates for the ATT}
In this section, we compare ATT estimates obtained using the TWFE method with ATT estimates obtained using the DRDID method. The TWFE estimate is obtained by running the linear regression \ref{eq:twfe} with an ordinary least squares (OLS) estimator, with the same covariates as the DRDID method.
The results for electric consumption are given in \Cref{tab:twfe_att_elec}, and the results for the gas consumption are given in \Cref{tab:twfe_att_gas}.

The results are similar to the ones obtained using the DRDID method (\Cref{tab:elec_results} and \Cref{tab:gas_results}). The main different between the two estimations are the estimated standard errors, which are 2 to 3 times larger for the TWFE method. We have no explanation for this difference.

\begin{table}[!ht]
    \centering
    \footnotesize
    \begin{tabular}{lrrrrr}
        \toprule
        retrofit measure    & ATT    & s.e. & LCB    & UCB   & p-value \\
        \midrule
        changing windows    & -2.99  & 2.28 & -7.18  & 1.40  & 0.179   \\
        comprehensive       & -9.16  & 4.68 & -16.95 & -0.63 & 0.036   \\
        heat pump           & -12.80 & 3.33 & -18.23 & -7.01 & <0.001  \\
        insulating attic    & -5.69  & 1.67 & -8.71  & -2.57 & <0.001  \\
        insulating basement & -1.64  & 5.71 & -11.79 & 9.69  & 0.766   \\
        insulating walls    & -7.22  & 3.12 & -12.64 & -1.46 & 0.015   \\
        \bottomrule
    \end{tabular}

    \caption{TWFE estimates for the ATT for electricity consumption, in \%. s.e. is the standard error, LCB is the lower bound of the 95\% confidence interval, and UCB is the upper bound.}
    \label{tab:twfe_att_elec}
\end{table}

\begin{table}[!ht]
    \centering
    \footnotesize
    \begin{tabular}{lrrrrr}
        \toprule
        retrofit measure    & ATT    & s.e.  & LCB    & UCB   & p-value \\
        \midrule
        changing windows    & -5.37  & 6.59  & -16.51 & 7.26  & 0.387   \\
        comprehensive       & -16.26 & 12.97 & -34.11 & 6.43  & 0.146   \\
        insulating attic    & -7.05  & 4.74  & -15.12 & 1.79  & 0.115   \\
        insulating basement & -0.56  & 15.77 & -25.52 & 32.75 & 0.969   \\
        insulating walls    & -16.41 & 7.92  & -28.03 & -2.92 & 0.019   \\
        \bottomrule
    \end{tabular}
    \caption{TWFE estimates for the ATT for gas consumption, in \%. s.e. is the standard error, LCB is the lower bound of the 95\% confidence interval, and UCB is the upper bound.}
    \label{tab:twfe_att_gas}
\end{table}

\end{document}